# Open Source Software Opportunities and Risks

John Sherlock, Manoj Muniswamaiah, Lauren Clarke, Shawn Cicoria
Seidenberg School of Computer Science and Information Systems
Pace University
White Plains, NY, US
{js20454w, mm42526w, lc18948w}@pace.edu, shawn@cicoria.com

*Abstract*-Open Source Software (OSS) history is traced to initial efforts in 1971 at Massachusetts Institute of Technology (MIT) Artificial Intelligence (AI) Lab, the initial goals of OSS around Free vs. Freedom, and its evolution and impact on commercial and custom applications. Through OSS history, much of the research and has been around contributors (suppliers) to OSS projects, the commercialization, and overall success of OSS as a development process. In conjunction with OSS growth, intellectual property issues and licensing issues still remain. The consumers of OSS, application architects, in developing commercial or internal applications based upon OSS should consider license risk as they compose their applications using Component Based Software Development (CBSD) approaches, either through source code, binary, or standard protocols such as HTTP.

## I. OPEN SOURCE SOFTWARE BACKGROUND

Open Source Software (OSS) terminology is traced to a community code sharing model at the Massachusetts Institute of Technology (MIT) Artificial Intelligence (AI) Lab in 1971[59]. In conjunction with the MIT AI community, the code sharing activities introduced the Free Software model with no restrictions on modification, reuse in other systems, or even acknowledgement of prior work [59].

The concept of Free with OSS has been linked with zero cost [59]; however, the initial intent was Freedom, where price was irrelevant. That Freedom was intended to include [59]:

• You have the freedom to run the program, for any purpose.

• You have the freedom to modify the program to suit your needs. (To make this freedom effective in practice, you must have access to the source code, since making changes in a program without having the source code is exceedingly difficult.)

• You have the freedom to redistribute copies, either gratis or for a fee.

• You have the freedom to distribute modified versions of the program, so that the community can benefit from your improvements.

In 1984, the free GNU operating system [60] was released with grounding in the model of Freedom. Later, Linux was introduced in 1991, and the social aspects of bazaar type programming [53], where the contributions of many and egoless [53] programmers, come together to collectively create software.

### A. Evolution of OSS

OSS software has become an integral part of the ecosystem for software in both free and commercial models. The commercial size of the market was estimated at $1.8 billion in 2006[5] and was projected to grow to $5.8 billion in 2011[5], based upon total software market size in 2011 of $245 billion [51].

The success of OSS has been measured by various means. One approach is based upon the total number projects hosted in OSS repositories such as SourceForge, github, and Google Code; other measures of perceived success include lines of source code, number of committers, downloads and user satisfaction[36]. Several maturity models have been developed to aid evaluation by consumers and help adoption [50]. Regardless of the measure, OSS as a model of development, regardless of the commercial aspects, has become integral part of the total market with some presence in many major software products [15].

Even the attitude of key antagonists to the OSS efforts changed their view of OSS over time. Large proprietary commercial software vendors such as IBM shifted to software and services during the 1990s as the OSS movement was accelerating [44]. IBM also formed an alliance with Red Hat [44], a key Linux distribution vendor created in 1995[24]. Even Microsoft, considered a key opponent to the OSS movement [20, 37] has changed its perspective. Microsoft was initially hostile[20]; however, has moved towards a more open model, as Microsoft has established its own OSS License types - Microsoft Public License (MS-PL) and the Microsoft Reciprocal License (MS-RL)[42], approved in 2007 by the OSI[65], along with OSS code repositories such as CodePlex[40], and frameworks such as ASP.NET MVC[38,39].

The confusion of Free (vs. Freedom) led to the organization of the Open Source Initiative (OSI) in 1998[59] with the intent of facilitating collaborative development for commercial purposes [63] and effectively a development methodology [53, 60] for software creation. The effectiveness and success of the OSS model continues to be debated [10] from various perspectives including the Community OSS model and Commercial OSS [13, 30, 36, 54].

OSS terminology continues to cause confusion in terms of what the consumer of that software can do with the source code, object code, libraries, and application programming interfaces (API)[14,25]. How the software can be used is coupled to intellectual property law [12] and the various OSS license types used by OSS. There has been a proliferation [25] of OSS license types, with hundreds of OSS license types [33] that exist for authors; although, GPL remains the leading choice – applied on more than 50% of OSS projects [33]. The GNU Lesser-GPL (LGPL) is a distant second with just under 10% [33]. While GPL remains the leading OSS licensing model, many in the OSS community still are conflicted on free vs. freedom, with the OSI community summarily defining free as "not using the GNU General Public License [60]."

### B. Benefit of OSS

The commercial, economic, and social viewpoints for those that participate in creation (committers [49, 54, 56]), management, or commercialization of OSS software continues to be debated [36]. The consumers of OSS, for commercial packages or internal organizational applications, are challenged with different set of decision points that include productivity, costs, market timing, intellectual property, license, legal, and liability concerns.

Building software applications today generally is a compositional approach [6, 55]. Application implementations are comprised of a mix of parts, in a component-based software development approach [45, 52] that are sourced from other work – sometimes internal to an organization, sometimes OSS. This model of reuse can be in source code, where the package or source is compiled alongside an application and embedded, through binary (reference), even across standard protocols such as HTTP.

The building block approach [9] to application development intent is to decrease costs and time-to-market of building an application [3]. Debate as to the quality of OSS components contributing to overall application quality still exists [30, 44, 66] and should be a consideration in whether or not to choose OSS.

## II. OSS LICENSE

OSS licenses exist to permit and encourage the non-exclusive development, improvement and distribution of the licensed software works. A fundamental purpose of OSS licensing is to deny anyone the right to exclusively exploit a work [35]. Sometimes referred to as 'free' software, the work product licensed is considered to be freely modifiable and freely distributed. This approach to software development promotes [35, 53]:

● *Innovation:* Programmers contributing excellent work products, adding value to existing work product.

● *Reliability:* Knowledgeable users collaborating on testing and fixing work product.

● *Longevity:* Work product that would have otherwise reached its 'end of life' continues revived, adapted or rewritten as a new work.

The basic principles of OSS licensing (as per the Open Source Definition, as propounded by the Open Source Initiative) are [35, 60]:

● *Free Distribution:* Open Source licenses must permit non-exclusive commercial exploitation of the licensed work

● *Source Code:* Must make available the work's source code. Deliberately obfuscated source code (and intermediate forms) is not allowed.

● *Derived Works:* Must permit the creation of derivative works from the work itself. Licensee is not necessarily barred from 'going closed' (the work being incorporated into proprietary code). In the case of derivative work, the license may require a different name and/or version number.

There are other restrictions such as remaining technology-neutral and interface-neutral. OSS licenses typically grant the right to copy, modify, and distribute source and binary code, while proprietary licenses may grant only the right to possess one or a limited number of binary copies. OSS licenses typically impose an obligation to retain copyright and license notices unmodified [2].

When licensing the free use and distribution of software works, included also must be the source code. The licensee must be free to make modifications to the licensed works, albeit usually with certain conditions, limitations and obligations. These stipulations do become more onerous depending on the type of Open Source license used by the Licensor.

Warranty disclaimers are also common in Open Source licenses (protecting the Licensor against potential liabilities) [35]:

● These disclaimers can sometimes be nullified (based on a contrary, previously unknown, representation or agreement).

● Certain state and federal laws may limit effectiveness of these disclaimers.

All conditions (included for the protection of the Licensor) need to be carefully reviewed by and Licensee before accepting (for their protection).

### A. OSS Reuse Patterns

When using Component Based Software Development (CBSD) approaches, it's important to understand how components are reused in applications [9] and potential implication for usage rights. Many commercial software vendors use informal or non-objective evaluation models when choosing to incorporate OSS [31].

We will briefly describe 3 high level patterns of on-premise reuse and 1 partner model for coupling of components in application development. The coupling choice can have an impact on adherence to licensing constraints OSS authors publish under [1]. For example, under the GNU GPL object code (non-source) is specifically identified [18]. With Open Architectures (OA) [1] and Open API [1, 48], reuse is over standard protocols, such as HTTP, in consumption of capabilities from external OSS or service providers.

OSS libraries aren't necessarily modified for any of these approaches. OSS can be used as source code (unmodified), object libraries (binary references), or over standard or proprietary protocols such as HTTP or Microsoft Exchange Server [41]. When using OSS as direct source code, coupling to the OSS capabilities is direct.

The example patterns of reuse are (see Appendix A – CBSD Reuse Patterns – for more detail):

- Direct or Compiled Source Code (on-premise)

- Binary reference, static or dynamic (on-premise)

- Inter-process or distributed (on-premise)

- Inter-process or distributed (partner / service provider)

The first 3 represent composition of an application and use of OSS components that are consumed directly at a location under the control of the consumer's data center. The 4th pattern represents consumption through a partner Open API. The Google Map API, Microsoft Bing API, and similar service provider models, generally located outside of the consumer's data center, are examples. For simplicity, we will exclude models of Open API use for appliance type installations (e.g. Windows Azure Appliance [43], Google Search Appliance [26]).

### B. Other Patterns Considerations

When considering the combinations of available architectural design decisions (software components and component relationships) and available OSS licenses, one is faced with considering what has become known as "open source legality patterns" [29]. The generic goal of patterns is to define a recurring problem in a context, identify a solution to the problem (typically in an existing system), and document consequences. Once defined, the respective OS component is identified (if applicable) and that software rendered covered by the respective OS license. The specific goal of open source legality patterns is to identify and manage the way different software components interact to ensure that all licenses of open source components are complied with [29]. There have been identified various legality patterns types which one should consider in addressing this area of licensing risk [29]:

• Interaction: related to the client-side user interface and data communication

• Isolation: server-side functionality

• Licensing: how package (application) components should be licensed/relicensed (separately or combined, tiered, etc.)

In addition to legality patterns, other type of patterns have been defined and have become a central part of contemporary software engineering [29]:

• Architectural [8]
• Design [32]

• Analysis [17]

Considering the problems introduced when using various incompatible OSS licensed software components, combined with the corresponding legal challenges of even understanding whether there are license conflicts given the complexities of the documents, the concept of legality patterns is an attempt to logically separate different software components in order to eliminate the conflict (i.e., eliminate the viral effect of strong copyleft restrictions). This viral effect is considered harmful by some companies developing proprietary software that interacts with OS components [29]. Under the copyright laws of the United States, copyright is automatically attached to every novel expression of an idea whether through text, sounds, or imagery [35]. It is only the creator of the work that inherits the right to create derivative works from this original copyrighted creation. The expression of the idea (i.e., how to solve a given problem, or how to render the results on a screen, or how to combine certain bit and bytes together to form a solution) is the basis of software coding. The underlying substance of the idea is what a patent would serve to protect. This privilege of copyright is certainly applicable to the works under consideration for this discussion of OSS.

The rights assigned under copyright law have a very long life: the life of the creator plus 70 years, or in the case or works made for-hire or by creators who are not identified, 95 years from the date of publication or 120 years from the date of creation, whichever is shorter [35].

In order to succeed in a claim for infringement of copyright in computer programs, a copyright holder has to show [35]:

• That copyright is capable of subsisting and in fact subsists in the work at issue

• That he/she is the owner of the copyright

• That acts have been carried out within the exclusive rights of the right holders

• That those acts amount to infringement

In consideration of the constructs of OSS license there are two limitations which influence the liability and enforceability of OSS licenses: the doctrines of work-for-hire and fair-use[35]:

• *Work for Hire:* Related to works generated by en employee during the employ of another. Work-for-hire works are still subject to copyright, but the rights belong to the employer

• *Fair Use:* Related the right of a person to make certain limited uses of copyrighted materials for the purpose of commenting, criticizing, reporting or teaching

Two other copyright-related limitations should be mentioned as relevant to any discussion on the non-infringing nature of OSS initiatives [35]:

• *Transformative Derivative Work*: Work based on copyrighted work which is so fundamentally altered from the original that it is considered a new work.

• *Time:* After the legitimate expiration of the copyrighted work, that work goes into the public domain, free for anyone to commercially exploit.

## C. Copyleft

Copyleft is an inheritance requirement to pass on the GPL's terms to other software that contains or is derived from the initially used GPL software [2]. Copyleft says that anyone who distributes the software, with or without changes, must pass along the freedom to further copy and change it. It would therefore be illegal to distribute the improved version except as free [19].

As copyright is a right to exclude others, copyleft is a requirement that licensees be included in development, distribution, and source code access rights, but always under the copyleft license. Copyleft licenses exclude other inconsistent licenses, which renders them "incompatible" with commercial licenses and some, but not all, OSS licenses. To determine whether two OSS licenses are compatible, you need to read and compare both. Version 2 of the GPL is incompatible, for example, with the Apache Public License, a copyleft agreement that covers the popular Apache server [34].

Incompatibility is not a problem, however, as long as you keep the two programs separate, even if they both operate in the same computer. If, ignoring the GPL terms, you distribute the resulting combined program under a proprietary license, then the included GPL-licensed code would be unlicensed (because you distributed it under an incompatible license), and your unlicensed distribution therefore would infringe the copyright on the GPL-licensed code. The copyright owner of the GPL-licensed code presumably could sue you for copyright infringement.

Copyleft clauses do not affect programs that are clearly separate. For example, IBMs Websphere, a Web portal manager program, runs on GPL-licensed Linux and may be shipped with Linux, but it remains fully an IBM proprietary program.

## D. License Selection

A significant consideration by an OSS Licensor is which license type to use (see Appendices B and C for a comparison of license types). A major contributor to this decision process is whether the software project involved development of a new product, or if the project (and/or the work product) has been inherited ('handed down') from someone else: i.e., new development or patch work. Inherited project (with their associated OSS licenses) sometimes may involve various administrative and legal difficulties. The new project leader (the potential licensor of the derivative work) would like likely need to secure the consent of every programmer who had contributed to the project under any previous license. After all, they made their contributions with the understanding that what they contributed would be licensed under the license applicable to that original project [35].

Still, decisions for the license of the derivative work may need to be determined based on the license type of the original work. By scanning online services (such as SourceForge.net), one can see whether sufficiently similar work has been conducted in the past or is currently under way. In many cases, the licensor's options are constrained by choices made by the predecessor [35].

The most important decision will be whether to use the GPL template or a less restrictive type. GPL is set up to encourage open development models, yet discourage reliance on software not developed under open development (including all proprietary software). There is a strong incentive for programmers to follow through and continue the GPL licensing for their derivative works. An argument for using a less restrictive license type would be the promotion of other development models (not just open development) and the inclusion of proprietary code ('closed' code) into the ultimate solution.

## E. License Restrictiveness

License restrictiveness can be stratified into three areas [67]:

- Strong copyleft (highly restrictive, require derivative works to inherent the license)

- Weak copyleft (less restrictive, only require derivative work to be licensed similarly)

- Non-copyleft (non-restrictive, derivative works are not required to inherent the license)

License restrictions are found to be positively associated with OSS project survival in the initial stages of the project (when team members are first collaborating on the product), yet is found to have no impact on project survival at the growth stage (after the product has been released and has established a usage track record).

Studies on open source software (OSS) have shown that the license under which an OSS is released has an impact on the success or failure of the software. Sen, Subramaniam and Nelson state that:

The optimal license choice for original OSS is a function of the preferred license of the original OSS's developer(s), the effort that goes into developing the original OSS and any derivative software base on this OSS, and the value to the other developers of the original OSS and any derivative OSS. In subsequent discussions we assume that the OSS being developed has high value for the developers working on the OSS project. This assumption is based on the fact that most OSS developers work on a completely voluntary, non-contractual, non-commissioned basis and suggests that motivation plays a significant role in their behavior. The key motivational factors identified in existing literature include the solving of information technology problems in day-to-day working, and reputation and recognition by peers. In light of these motivational factors we can safely assume that the OSS being developed has a high intrinsic value for the developers [57].

This study did conduct a survey of the Sourceforge.net database which contains information on more than 200,000 software projects. For the purpose of this study we considered only those projects for which complete information was available, and which had been registered between January 1999 and December 2005. The number of such projects was 10,094. Approximately 66% of these were licensed as strong-copyleft, about 16% as weak-copyleft and the rest as non-copyleft [57].

## F. Cross Licensing Options

How does a programmer combine the elements from two or more programs (each under separate licenses, possibly of different types, possibly incompatible) into a new program, and not violate the terms of either original license? The general advice to the programmer is to proceed cautiously [35]. The specific advice is to execute a cross license making the program available under a license other than that which the program was originally provided under. This is considered in Section 10 of the GPL [35] allowing the licensee to further cross-license (license the same (unaltered) original work under another license). There are sometimes limitations to cross-licensing; the licensee must check the original license carefully (or request the assistance of an attorney).

### G. Forking (Splitting Projects)
Forking occurs when software projects split. Sometimes this is unavoidable and even necessary. Forking on early stage project have be handled with relate easy; forking on mature projects are properly feared. It is not unreasonable to look to licenses to prevent or at least to decrease the probability of forking. The GPL limits the likelihood of forking by prohibiting non-open development models for projects that incorporate GPL-licensed code [35]. After a fork on a GPL project, each leg of the project remains free to draw on the work of the other leg(s). Is intended for this process to hasten the closing of the fork and permit the reunification of the forked project. It does not always turn out this way as the nature of open development is conducive to forking.

## III. LICENSING RISK
OSS is an attractive option for software development efforts. However, with OSS there are risks. There are a number of risk-related issues which the licensee should be considering in the selection of an OSS license type:

● The possibility of being exposed to copyright and/or patent complaints and/or infringements. It can be very difficult to trace back originals of the preceding works, therefore difficult to identify the original licenses.

● Failure to comply with license terms will results in the automatic termination of the license; if the programmer continues to use the respective OSS, it becomes copyright infringement and the guilty party may be prosecuted [28].

● Overlapping (and conflicting) OSS licenses. Some may not be combined under any circumstances (cross-licensing not permitted).

● OSS licenses are perpetual; once you accept, there is no time limit on Terms of Use.

● Having a high regard for 'openness', OSS licenses strive to have all software using their source to also be publicly available. As such, most OSS licenses stipulate that one cannot license patents exclusively or under special terms with one company, while blocking others. The same terms must be given to all who license the software.

Risks can be grouped into three primary OSS risk areas: 1) strategic, 2) operational, and 3) legal.

### A. Strategic Risk
Strategic risks include the ability to customize and maintain the code, compatibility and interoperability, systems integration and support and total cost of ownership.

### a. Ability to Customize
Companies will customize OSS for their own uses. They should test to ensure the integrity of systems and data carefully consider their technical and legal ability to modify and maintain code [16]. They should also ensure that controls are in place to protect against patent and copyright infringement [16].

### b. Compatibility and Interoperability
Since OSS is generally written to open standards it is usually more interoperable than proprietary software. However, the interoperability of OSS programs may not be formally [16]. Therefore companies using OSS should ensure that it meets their needs for compatibility and interoperability. Additional staff or vendors with an expertise in software integration may need to be hired and/or consulted.

### c. Systems Integration and Support
OSS can be acquired and implemented with varying degrees of integration and support. If a systems integrator is used, they ensure compatibility for all OSS components. Conversely, if OSS is obtained from development projects, integration is done in-house. Consideration should be given to the identification tracking, evaluation, modification, installation and maintenance of the software [16].

### d. Total Cost of Ownership
Both direct and indirect costs should be taken in to consideration when evaluating the total cost of ownership of OSS. Direct costs include hardware, licensing and maintenance. Indirect costs may include additional training for staff and change management. More resources may be responsible for identifying analyzing, installing, upgrading and patching the OSS. Indirect costs that may not be considered for OSS are costs for code reviews, documentation and contingency planning [16].

### B. Operational Risks
Operational risks include code integrity, sufficiency of documentation, contingency planning and external support.

### a. Code Integrity
Since the OSS is widely available and can be distributed by anyone verification of code integrity is important. Companies need to adopt standards and put in place procedures to ensure they are acquiring source code from trustworthy sources and verify the code once it is received. This should also apply to patches and updates.

### b. Documentation
The documentation that comes with OSS is usually inadequate and less comprehensive than documentation that would accompany proprietary software. Companies should, upon considering an OSS set of code, have a minimum set of documentation requirements and also have in place a staff to further expand on the documentation.

### c. Contingency Planning
The continued viability of OSS is largely dependent on the OSS community and third-party vendors [16]. But companies should develop a contingency plan if the software ends up not being developed further and support goes away. Also, if litigation is imposed around a set of code companies may want to abandon further use of it.

### d. Support
External support for OSS has become more robust and users are no longer as dependent on informal support such as development communities and Internet mailing lists [16]. Since OSS has gained in popularity, there are many options for support such as Value Added Resellers (VARs) and independent developers. Companies need to be prudent about finding ongoing support for the products they incorporate into their systems.

### C. Legal Risks
OSS and Licensing Impedance for products composed of OSS and non-OSS. There represents a licensing mismatch [22] that can create serious legal issues for enterprises, commercial software vendors, and event software services venders.
Two key legal risks of using OSS include licensing and copyright and patent infringement.

### a. Licensing
One key to avoiding the risks of OSS is to have a good understanding of the license types.
There are between fifty and seventy different types of Open Source licenses each with different rights and restrictions. Since there are so many different license types, having a good understanding of the various license types, from a legal counsel perspective is advisable. For the most part, OSS licenses permit copying, distribution and modification of the source code with no warranty or indemnification [16]. It is recommended to have legal counsel available to review the licensing options based on the company's strategy is for its intended use.

### b. Infringement
Noncompliant use of OSS put companies at a higher risk of being sued for patent or copyright infringement [69]. This is due to the fact that OSS is developed by individuals in an open environment where code is shared and developed by numerous individuals. The code sharing increases the possibility that proprietary code may be inserted in the OSS somewhere in the development process [16] Other Licensing Considerations

### D. Software Origins
A licensee never really knows the provenance of any software it obtains. While this risk exists for all software, the risk is clearly higher for collaboratively developed OSS. (This risk would likely be much lower for OSS created by a single company.) As discussed above, some very important OSS products, such as Linux and Apache, are the result of a process in which hundreds of individuals have contributed code. For those products, there is no way to be sure that each contributor actually had the rights under copyright law to make the contribution. Therefore, collaboratively developed OSS products carry an inherent risk that they might include code included without permission and in violation of some unknown copyright holder's rights [34].

### E. An Informed Decision
Software development efforts exacerbate risk when not following key mitigation approaches [11]:

• Code inspection process and guidelines

• Management infrastructure to support the process and guidelines

• Comprehensive knowledge resource for license compliance

• Active mindset on OSS

• Guarantees on quality of some OSS

It has been estimated that 25% of software developers had never received any form of training or information on the topic of OSS licensing [58]. In this same survey, 50% of developers surveyed deemed ad hoc reuse at least 'somewhat important' for their own work [58]. This result differs from the prevailing assumption of many firms that their code base does not contain Internet code[58].To assist in the analysis of a software application (and its respective components, connectors and sub-systems) in order to determine what inherent rights, obligations and constraints exist, automated software tolls can be used. ArchStudio4 is an example of such a tool [2]. The tool does assume a certain level of codifying of parameters about the various application components and about the respective software licenses. These annotated software architectural descriptions can be prescriptively analyzed at design-time, at build-time or at run-time [2].

## IV. FURTHER RESEARCH
In review of the literature, there are several areas that are of interest for follow up research, and in some emerging areas, a lack of research.

### A. License Validation Process and Tools
There are examples of patterns [23] and tools [22] for license mismatch and validation within a code base and applications. In addition, other OSS and commercial tools exist in aiding the overall process for repositories of OSS and tooling for automated validation [46]. While industrialization is occurring, the true efficacy in these tools is still not independently evident and represents an area of research opportunity.

### B. Open API and Cloud Hosted Services and Applications
Cloud based services offered from commercial, non-commercial, and public entities, there can also be republishing of services as Open API [21]. These Open API's can also be described as Open Architecture (OA) [1]. These OA applications, products, and hosted services are composed of various components each having their own license. They in turn could also be composed of further sub-components, again, with their own license – and subject to a licensing mismatch. Most research areas focused on Commercial Off-The-Shelf software (COTS), but not the public API of cloud services and applications many solutions are now built upon. An example of the complexity and seemingly contradiction of expectations is provided for in the BugZilla[7] application which while using a component that is licensed under the GNU GPL[18], considered the most viral[68] still results in a commercially viable product licensed under a seemingly incompatible license model[22]. Additionally, jQuery distribution [61] through Content Delivery Networks (CDN) are areas that are not fully researched.

## V. APPENDICES

### A. *Appendix A – CBSD Reuse Patterns*

In the following diagrams we use the following terminology:

- Premise – represents the location or data center of the consumer, or facility under control of the consumer

- Host – represents the process containing the application

- Application – represents the solution that is composing or leveraging the OSS for capabilities required

- Partner – represents the external organization or service provider exposing an Open API for consumption by applications

### Direct or Compiled Source Code (on-premise)

This pattern is a compiled or directly referenced within the overall application. Generally, the OSS components, source code is considered tightly coupled with the application and potentially distributed as part of the application.

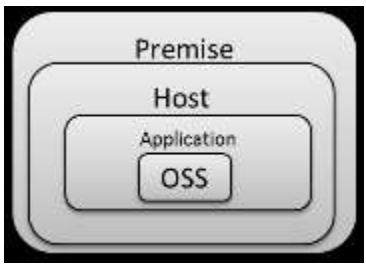

**Figure 1 - Direct or Compiled Source Code (on-premise)**

### Binary reference, static or dynamic (on-premise)

This pattern of reuse, OSS object-code, library, or, assembly (terminology varies) is referenced unmodified by an application. The OSS component is provided in a precompiled (published) distribution with a black-box reuse model. The published API of the component is directly used by an application, but not necessarily distributed directly with an application.

There are examples of source code being leveraged in this manner without compilation, further complicating the pattern discussion. As an example, the jQuery[62] library is distributed under the MIT License[64] in source code and directly referenced and used within applications without direct application producer controlled distribution. These OSS components are consumed directly by application users through HTML JavaScript SRC tags direct from Content Distribution Networks (CDN) provided by Google or Microsoft [61].

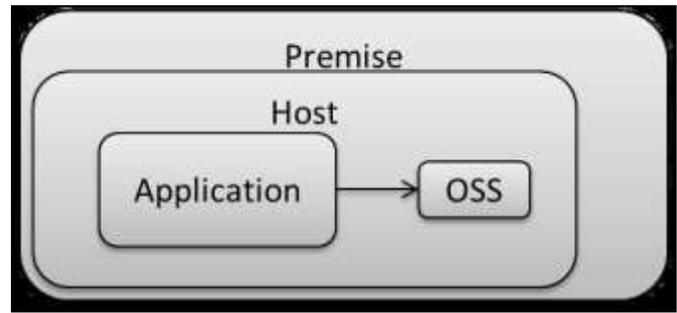

**Figure 2 - Binary reference, static or dynamic (on-premise)**

### Inter-process or distributed (on-premise)

This pattern leverages inter-process communication or protocols either within a host, or across a host. The key aspect is that the OSS components are accessed across different processes, but are physically deployed within the consumer's span of control or data center.

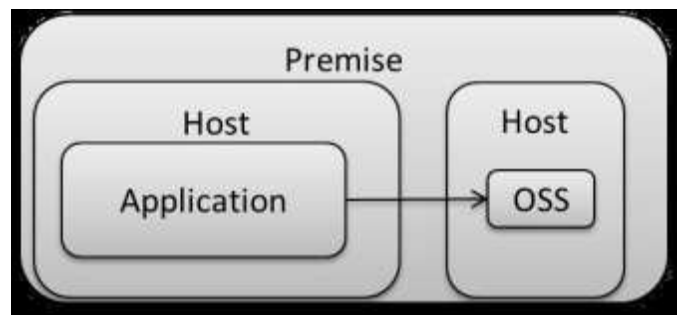

**Figure 3 - Inter-process or distributed (on-premise)**

### Inter-process or distributed (partner / service provider)

This pattern is an approach used by applications to consume an Open API published by service providers. This could be commercial or non-commercial. Generally, Open API publishers have their own license models related to liability and service level agreements [4, 27].

While many of the Open API service providers use standard protocols, it is not a requirement [41].

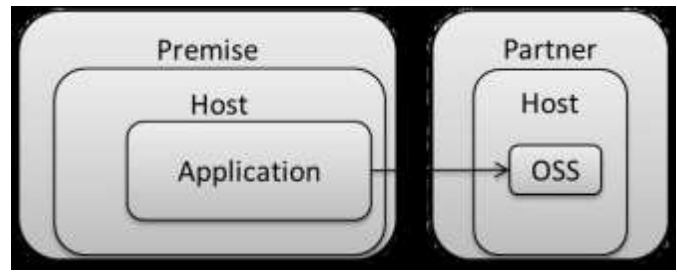

**Figure 4 - Inter-process or distributed (partner / service provider)**

## B. Appendix B – Comparison of OSS License Types

| License Type | Highlights of the License Type |
|---|---|
| The MIT (or X) License | Right to use, copy, modify, distribute and sublicense. Must include copyright notice in derivative works, not necessarily in the original work as used. Software is provided 'as is', no warranty of any kind. |
| The BSD License | Early versions (prior to 1999) required an acknowledgement notice in the derivative work. Names of work contributors may not be used to endorse or promote derivative work. Variations included FreeBSD, NetBSD and OpenBSD. |
| The Apache License, v1.1 and v2.0 • The Apache License v1.1 • The Apache License v2.0 | *v1.1:* Documentation must include acknowledgement. No mention of 'Apache' in derivative work. *v2.0:* Long definitions sections added Patent rights addressed. Licensor (contributor) grants copyright and patent licenses, but not trademarks. Any notices from original work must be retained in the derivative work. |
| The Academic Free License | Similar to Apache v1.1 with some clarifications and some further restrictions. |
| GNU General Public License GNU *Lesser* General Public License | Requires derivative works be distributed under the GPL license (the idea of 'copyleft'). Further restrictions on licensing for derivative work. Derivative works require a distinguishing version number Included instructions on how to implement the licensor for derivative work *Lesser:* Addresses certain classes of programs (e.g. subroutine libraries); Slightly less restrictive on conditions of use. |
| The Mozilla Public License 1.1 (MPL 1.1) (originally Netscape Public License (NPL)) | Basically a hybrid (a 'middle ground') between GPL and BSD. Permits the use of the 'Covered Code' in Larger Works" Accommodations for Contributor APIs. |
| The Q Public License (QPL) | Sometimes cross-licensed with the GPL. Not a very commonly used license type. |
| Artistic License (Perl) | Also typically cross-licensed with the GPL. Not a popular license type because some license terms are vague and confusing |
| Creative Commons Licenses • Attribution-ShareAlike *Version 1.0* • Attribution-ShareAlike *Version 2.0* | Not original intended for the software industry (rather music, web site content, and film) encouraging creators to place their work in the public domain. Does not distinguish between commercial and non-commercial works. Addresses 'fair use' |

## Appendix C – Comparison of OSS License Types

| License Types[47] License | Ownership | Virality | Inheritance |
|---|---|---|---|
| GPL | No | Yes | Yes |
| CeCILL | No | Yes | Yes |
| LGPL | No | Partial | Yes |
| BSD | Yes | No | No |
| Artistic | Yes | No | No |
| MIT | Yes | No | No |
| Apache v1.1 | Yes | No | No |
| Apache v2.0 | Yes | No | No |
| MPL v1.1 | No | No | Yes |
| Common Public License V1.1 | No | No | No |
| Academic Free License V2.1 | Yes | No | No |
| PHP License v3.0 | Yes | No | No |
| Open Software License v2.0 | No | No | No |
| Zope Public License v2.0 | Yes | No | No |
| Python SF License v2.0 | Yes | No | No |

**Ownership** – can the derived code become proprietary or must it remain free?

**Virality** – is another module linked to the source code inevitably affected by the same license?

**Inheritance** – does the derived code inherit inevitably from the license or is it possible to apply additional restrictions to it?